\documentstyle[twoside]{article}

\catcode`\@=11
\long\def\@makefntext#1{
\protect\noindent \hbox to 3.2pt {\hskip-.9pt  
$^{{\eightrm\@thefnmark}}$\hfil}#1\hfill}		%CAN BE USED 

\def\@makefnmark{\hbox to 0pt{$^{\@thefnmark}$\hss}}	%ORIGINAL 
	
\def\ps@myheadings{\let\@mkboth\@gobbletwo
\def\@oddhead{\hbox{}
\rightmark\hfil\eightrm\thepage}   
\def\@oddfoot{}\def\@evenhead{\eightrm\thepage\hfil
\leftmark\hbox{}}\def\@evenfoot{}
\def\sectionmark##1{}\def\subsectionmark##1{}}

\oddsidemargin=\evensidemargin
\newcounter{sectionc}\newcounter{subsectionc}\newcounter{subsubsectionc}
\renewcommand{\section}[1] {\vspace{12pt}\addtocounter{sectionc}{1} 
\setcounter{subsectionc}{0}\setcounter{subsubsectionc}{0}\noindent 
	{\tenbf\thesectionc. #1}\par\vspace{5pt}}
\renewcommand{\subsection}[1] {\vspace{12pt}\addtocounter{subsectionc}{1} 
	\setcounter{subsubsectionc}{0}\noindent 
	{\bf\thesectionc.\thesubsectionc. {\kern1pt \bfit #1}}\par\vspace{5pt}}
\renewcommand{\subsubsection}[1] {\vspace{12pt}\addtocounter{subsubsectionc}{1}
	\noindent{\tenrm\thesectionc.\thesubsectionc.\thesubsubsectionc.
	{\kern1pt \tenit #1}}\par\vspace{5pt}}

\newcounter{appendixc}
\newcounter{subappendixc}[appendixc]
\newcounter{subsubappendixc}[subappendixc]
\renewcommand{\thesubappendixc}{\Alph{appendixc}.\arabic{subappendixc}}
\renewcommand{\thesubsubappendixc}
	{\Alph{appendixc}.\arabic{subappendixc}.\arabic{subsubappendixc}}

\renewcommand{\appendix}[1] {\vspace{12pt}
        \refstepcounter{appendixc}
        \setcounter{figure}{0}
        \setcounter{table}{0}
        \setcounter{lemma}{0}
        \setcounter{theorem}{0}
        \setcounter{corollary}{0}
        \setcounter{definition}{0}
        \setcounter{equation}{0}
        \renewcommand{\thefigure}{\Alph{appendixc}.\arabic{figure}}
        \renewcommand{\thetable}{\Alph{appendixc}.\arabic{table}}
        \renewcommand{\theappendixc}{\Alph{appendixc}}
        \renewcommand{\thelemma}{\Alph{appendixc}.\arabic{lemma}}
        \renewcommand{\thetheorem}{\Alph{appendixc}.\arabic{theorem}}
        \renewcommand{\thedefinition}{\Alph{appendixc}.\arabic{definition}}
        \renewcommand{\thecorollary}{\Alph{appendixc}.\arabic{corollary}}
        \renewcommand{\theequation}{\Alph{appendixc}.\arabic{equation}}
        \noindent{\tenbf Appendix \theappendixc #1}\par\vspace{5pt}}
\newcommand{\subappendix}[1] {\vspace{12pt}
        \refstepcounter{subappendixc}
        \noindent{\bf Appendix \thesubappendixc. {\kern1pt \bfit #1}}
	\par\vspace{5pt}}
\newcommand{\subsubappendix}[1] {\vspace{12pt}
        \refstepcounter{subsubappendixc}
        \noindent{\rm Appendix \thesubsubappendixc. {\kern1pt \tenit #1}}
	\par\vspace{5pt}}

\newcommand{\textlineskip}{\baselineskip=13pt}
\newcommand{\smalllineskip}{\baselineskip=10pt}

\def\eightcirc{
\begin{picture}(0,0)
\put(4.4,1.8){\circle{6.5}}
\end{picture}}
\def\eightcopyright{\eightcirc\kern2.7pt\hbox{\eightrm c}} 

\newcommand{\copyrightheading}[1]
	{\vspace*{-2.5cm}\smalllineskip{\flushleft
	{\footnotesize International Journal of Modern Physics A, #1}\\
	{\footnotesize $\eightcopyright$\, World Scientific Publishing
	 Company}\\
	 }}

\def\abstracts#1#2#3{{
	\centering{\begin{minipage}{4.5in}\baselineskip=10pt\footnotesize
	\parindent=0pt #1\par 
	\parindent=15pt #2\par
	\parindent=15pt #3
	\end{minipage}}\par}} 

\renewenvironment{thebibliography}[1]
	{\frenchspacing
	 \ninerm\baselineskip=11pt
	 \begin{list}{\arabic{enumi}.}
	{\usecounter{enumi}\setlength{\parsep}{0pt}
	 \setlength{\leftmargin 12.7pt}{\rightmargin 0pt} %FOR 1--9 ITEMS
	 \setlength{\itemsep}{0pt} \settowidth
	{\labelwidth}{#1.}\sloppy}}{\end{list}}

\newcounter{itemlistc}
\newcounter{romanlistc}
\newcounter{alphlistc}
\newcounter{arabiclistc}

\newcommand{\fcaption}[1]{
        \refstepcounter{figure}
        \setbox\@tempboxa = \hbox{\footnotesize Fig.~\thefigure. #1}
        \ifdim \wd\@tempboxa > 5in
           {\begin{center}
        \parbox{5in}{\footnotesize\smalllineskip Fig.~\thefigure. #1}
            \end{center}}
        \else
             {\begin{center}
             {\footnotesize Fig.~\thefigure. #1}
              \end{center}}
        \fi}

\newcommand{\tcaption}[1]{
        \refstepcounter{table}
        \setbox\@tempboxa = \hbox{\footnotesize Table~\thetable. #1}
        \ifdim \wd\@tempboxa > 5in
           {\begin{center}
        \parbox{5in}{\footnotesize\smalllineskip Table~\thetable. #1}
            \end{center}}
        \else
             {\begin{center}
             {\footnotesize Table~\thetable. #1}
              \end{center}}
        \fi}

\def\@citex[#1]#2{\if@filesw\immediate\write\@auxout
	{\string\citation{#2}}\fi
\def\@citea{}\@cite{\@for\@citeb:=#2\do
	{\@citea\def\@citea{,}\@ifundefined
	{b@\@citeb}{{\bf ?}\@warning
	{Citation `\@citeb' on page \thepage \space undefined}}
	{\csname b@\@citeb\endcsname}}}{#1}}

\newif\if@cghi
\def\cite{\@cghitrue\@ifnextchar [{\@tempswatrue
	\@citex}{\@tempswafalse\@citex[]}}
\def\citelow{\@cghifalse\@ifnextchar [{\@tempswatrue
	\@citex}{\@tempswafalse\@citex[]}}
\def\@cite#1#2{{$\null^{#1}$\if@tempswa\typeout
	{IJCGA warning: optional citation argument 
	ignored: `#2'} \fi}}

\def\pmb#1{\setbox0=\hbox{#1}
	\kern-.025em\copy0\kern-\wd0
	\kern.05em\copy0\kern-\wd0
	\kern-.025em\raise.0433em\box0}

\def\fnt#1#2{\footnotetext{\kern-.3em
	{$^{\mbox{\scriptsize #1}}$}{#2}}}

\def\fpage#1{\begingroup
\voffset=.3in
\thispagestyle{empty}\begin{table}[b]\centerline{\footnotesize #1}
	\end{table}\endgroup}

\def\runninghead#1#2{\pagestyle{myheadings}
\markboth{{\protect\footnotesize\it{\quad #1}}\hfill}
{\hfill{\protect\footnotesize\it{#2\quad}}}}
\font\tenrm=cmr10
\font\tenit=cmti10 
\font\tenbf=cmbx10
\font\bfit=cmbxti10 at 10pt
\font\ninerm=cmr9

\font\eightrm=cmr8

\def\qed{\hbox{${\vcenter{\vbox{			%HOLLOW SQUARE
   \hrule height 0.4pt\hbox{\vrule width 0.4pt height 6pt
   \kern5pt\vrule width 0.4pt}\hrule height 0.4pt}}}$}}

\begin{document}

\runninghead
{AdS/CFT Duality and Conformality for Non-Abelian Orbifold} 
{AdS/CFT Duality and Conformality for Non-Abelian Orbifold} 

\normalsize\textlineskip
\thispagestyle{empty}
\setcounter{page}{1}

\copyrightheading{}			%{Vol. 0, No. 0 (1993) 000--000}

\vspace*{0.88truein}

\fpage{1}
\centerline{\bf AdS/CFT DUALITY AND CONFORMALITY FOR}
\vspace*{0.035truein}
\centerline{\bf NON-ABELIAN ORBIFOLD}
\vspace*{0.37truein}
\centerline{\footnotesize PAUL H. FRAMPTON}
\vspace*{0.015truein}
\centerline{\footnotesize\it Institute of Field Physics, 
Department of Physics and Astronomy,
Chapel Hill, NC 27599-3255 }
\baselineskip=10pt
\vspace*{0.035truein}

\bigskip
\bigskip

\abstracts{
An outline of the conformality approach to the gauge hierarchy
is given including the use of non-abelian orbifolds 
to give unified models of the left-right type.}{}{}

%\textlineskip			%) USE THIS MEASUREMENT WHEN THERE IS
%\vspace*{12pt}			%) NO SECTION HEADING

\vspace*{1pt}\textlineskip	%) USE THIS MEASUREMENT WHEN THERE IS

\bigskip
\bigskip

One can identify four approaches to the hierarchy problem by different
new (untested) physics at a TeV scale:

\begin{itemize}

\item Supersymmetry. GUT unification works better with TeV
supersymmetry.

\item Technicolor. New strong dynamics at a TeV scale.

\item Large extra dimensions at a TeV scale.    

\item Conformality at a TeV scale.  

\end{itemize}

Here, I discuss the fourth approach. It is motivated physically
by the notion that at a TeV scale the standard model appears almost conformally
invariant in the sense that the masses of the particles, as well as the QCD and weak scales
appear almost vanishingly small. But the theory is {\it not} conformal invariant as it
stands because the couplings still run. The conformality idea
is to enrich the spectrum just so
that the couplings cease to run at an infra-red
fixed point of the renormalization group.

Nevertheless as will become clear there can still be a gauge coupling unification
at the TeV scale in a larger gauge group. The disparity of the
321 couplings then arises from the group
theory of the embedding in the larger group.

\bigskip

Using AdS/CFT duality, one arrives at a class of gauge field theories
of special recent interest. The simplest compactification of a ten-dimensional
superstring on a product of
an AdS space with a five-dimensional spherical manifold leads to
an ${\cal N} = 4~SU(N)$ supersymmetric gauge theory, well-known to be
conformally invariant\cite{mandelstam}. By replacing the manifold $S^5$
by an orbifold $S^5/\Gamma$ one arrives at less supersymmetries
corresponding to ${\cal N} = 2,~1 ~{\rm or}~ 0$ depending
on whether $\Gamma \subset SU(2), ~~ SU(3),
~~{\rm or} \not\subset SU(3)$ respectively, where $\Gamma$
is in all cases a subgroup of $SU(4) \sim SO(6)$
the isometry of the $S^5$ manifold.

It was conjectured in \cite{maldacena} that such $SU(N)$ gauge theories
are conformal in the $N \rightarrow \infty$ limit. In \cite{F1} it was conjectured that
at least a subset of the resultant nonsupersymmetric
${\cal N} = 0$ theories are conformal even for finite $N$. Some
first steps to check this idea were made in \cite{WS}.
Model-building based on abelian
$\Gamma$ was studied further in \cite{CV,F2,F3}, arriving in \cite{F3}
at an $SU(3)^7$ model based on $\Gamma = Z_7$
which has three families of chiral fermions,
a correct value for ${\rm sin}^2 \theta$ and a conformal scale $\sim 10$~~TeV.

The case of non-abelian orbifolds bases on non-abelian $\Gamma$ has
now been studied\cite{FK}. 
We have considered all non-abelian discrete groups of order $g < 32$. These
are described in detail in \cite{books,FK2}. There are exactly 45 such
non-abelian groups. Because the gauge group arrived at
by this construction\cite{CV}
is $\otimes_i SU(Nd_i)$ where $d_i$ are the dimensions of the
irreducible representations of $\Gamma$, one can expect to arrive
at models such as the Pati-Salam
$SU(4) \times SU(2) \times SU(2)$ type\cite{PS}
by choosing $N = 2$ and combining two singlets
and a doublet in the {\bf 4}
of $SU(4)$. Indeed we find that such an
accommodation of the standard model
is possible by using a non-abelian $\Gamma$.

The procedures for building a model within such a conformality approach are:
(1) Choose $\Gamma$; (2) Choose a proper
embedding $\Gamma \subset SU(4)$ by assigning
the components of the {\bf 4} of $SU(4)$ to irreps of $\Gamma$,
while at the same time ensuring that the {\bf 6} of $SU(4)$ is real;
(3) Choose $N$, in the gauge group $\otimes_i SU(Nd_i)$.

We choose $N = 2$ and aim
at the gauge group $SU(4) \times SU(2) \times SU(2)$.
To obtain chiral fermions, it is necessary\cite{CV}
that the {\bf 4} of $SU(4)$
be complex ${\bf 4} \neq {\bf 4}^*$. Actually this condition is not quite
sufficient to ensure chirality in the present case because
of the pseudoreality of $SU(2)$. We must ensure that the {\bf 4} is
not pseudoreal.

This last condition means that many of our 45 candidates for $\Gamma$
do not lead to chiral fermions. For example, $\Gamma = Q_{2n} \subset SU(2)$
has irreps of appropriate dimensionalities for our purpose
but it will not sustain chiral fermions under $SU(4)\times SU(2) \times SU(2)$
because these
irreps are all, like $SU(2)$, pseudoreal.\footnote{Note that
were we using $N \geq 3$
then a pseudoreal {\bf 4} would give chiral fermions.}
Applying the rule that {\bf 4} must be
neither real nor pseudoreal
leaves a total of only 19 possible non-abelian discrete groups of
order $g \leq 31$. The smallest group which avoids pseudoreality
has order $g = 16$ but gives only two families. The
technical details
of our systematic search iare in \cite{FK}.

Here we mention only the
simplest interesting non-abelian case
which has $g = 24$ and gives three chiral families in a
Pati-Salam-type model\cite{PS}.

The first group that can lead to exactly three families occurs at
order $g = 24$ and is $\Gamma = Z_3 \times Q$ where $Q (\equiv Q_4)$
is the group of unit quarternions which is the smallest dicyclic group $Q_{2n}$.

There are several potential models due to the different choices for the {\bf 4}
of $SU(4)$ but only the case {\bf 4} = $(1\alpha, 1^{'}, 2\alpha)$ leads to three families.

Since $Q \times Z_3$ is a direct product group, we can write the irreps as $r_i \otimes \alpha^{a}$
where $r_i$ is a $Q$ irrep and $\alpha^{a}$ is a $Z_3$ irrep. We write $Q$ irreps as $1,~1^{'},~1^{''},~
1^{'''},~2$ while the irreps of $Z_3$ are all singlets which we call $1,~\alpha,|\alpha^{-1}$.
Thus $Q \times Z_3$ has twelve irreps in all and the gauge group will be of Pati-Salam type for $N = 2$.

If we wish to break all supersymmetry, the {\bf 4} may not contain a singlet of $\Gamma$.
Due to permutational symmetry among the singlets
it is sufficiently general to choose {\bf 4} = $(1\alpha^{a_1},~1^{'}\alpha^{a_2},~2\alpha^{a_3})$
with $a_1 \neq 0$.

To fix the $a_i$ we note that the scalar sector of the theory which is generated by the
{\bf 6} of $SU(4)$ can be used as a constraint since the {\bf 6} is
required to be real. This leads to
$a_1 + a_2 = - 2a_3 ({\rm mod}~3)$. Up to permutaions in the chiral fermion sector the most
general choice is $a_1 = a_3= +1$ and $a_2 = 0$. Hence our choice of embedding is
\begin{equation}
{\bf 4} = (1\alpha,~1^{'},~2\alpha)
\label{embed}
\end{equation}
with
\begin{equation}
{\bf 6} = (1^{'}\alpha,~2\alpha,~2\alpha^{-1},~1^{'}\alpha^{-1})
\label{six}
\end{equation}
which is real as required.

We are now in a position to summarize the particle content of the theory. The fermions are given by
\begin{equation}
\sum_I~{\bf 4}\times R_I
\label{fermions}
\end{equation}
where the $R_I$ are all the irreps of $\Gamma = Q \times Z_3$.

The scalars are given by
\begin{equation}
\sum_I~{\bf 6}\times R_I
\label{scalars}
\end{equation}

\bigskip

As described in more detail in \cite{FK}
the scalars are suffcient to break the starting
gauge symmetry $SU(4)^3\times SU(2)^{12}$ to the
required 4-2-2 left-right structure, and with precisely
three chiral families in 16-plets.

\end{document}